*Review*

# Kinetics and Thermodynamics of Membrane Protein Folding

**Ernesto A. Roman and F. Luis González Flecha \***

Laboratory of Molecular Biophysics, Institute of Biochemistry and Biophysical Chemistry, University of Buenos Aires-CONICET, Buenos Aires 1113, Argentina; E-Mail: ernest.roman@gmail.com

**\*** Author to whom correspondence should be addressed; E-Mail: lgf@qb.ffyb.uba.ar; Tel.: +54-11-4964-8289; Fax: +54-11-4962-5457.

**Abstract:** Understanding protein folding has been one of the great challenges in biochemistry and molecular biophysics. Over the past 50 years, many thermodynamic and kinetic studies have been performed addressing the stability of globular proteins. In comparison, advances in the membrane protein folding field lag far behind. Although membrane proteins constitute about a third of the proteins encoded in known genomes, stability studies on membrane proteins have been impaired due to experimental limitations. Furthermore, no systematic experimental strategies are available for folding these biomolecules *in vitro*. Common denaturing agents such as chaotropes usually do not work on helical membrane proteins, and ionic detergents have been successful denaturants only in few cases. Refolding a membrane protein seems to be a craftsman work, which is relatively straightforward for transmembrane β-barrel proteins but challenging for α-helical membrane proteins. Additional complexities emerge in multidomain membrane proteins, data interpretation being one of the most critical. In this review, we will describe some recent efforts in understanding the folding mechanism of membrane proteins that have been reversibly refolded allowing both thermodynamic and kinetic analysis. This information will be discussed in the context of current paradigms in the protein folding field.

**Keywords:** membrane proteins; thermodynamic stability; urea; guanidine hydrochloride; sodium dodecyl sulfate

## 1. Introduction

Natural proteins are heteropolymers composed of covalently bonded amino acid residues. They are ubiquitous and essential for living organisms. Most of them adopt a well-defined set of three-dimensional structures in solution, denoted as the folded or native ensemble. It is strongly established that protein structure and function are related [1], however, the discovery of functionally relevant "intrinsically disordered proteins" [2–4] and intrinsically disordered regions in structured proteins [5,6] opens new paradigms in structure-function relationships.

Two main groups of proteins can be recognized: the water-soluble and the membrane-related ones. The latter is a heterogeneous group that includes transporters, receptors and channels, sharing the characteristic of being inserted into biological membranes. Membrane proteins represent about 25%–30% of total proteins codified in known genomes [7,8] and they constitute the target of about 70% of current drugs [9]. Whereas globular proteins are organized around a hydrophobic core surrounded by a water accessible surface [1], membrane proteins have some of their more hydrophobic residues in the transmembrane surface interacting with phospholipids [10]. Moreover, as some of these proteins are involved in transport across membranes, the core of the transmembrane region may be hydrophilic, allowing the interaction of these regions with water-soluble ligands [11].

Pioneering work of Anfinsen *et al*. [12] has demonstrated that all the information required for a protein to fold is contained in its amino acid sequence. Since that work, folding of globular proteins, has been the subject of many experimental and theoretical studies, and as research methodologies are continuously improved, novel results are constantly found leading to new insights and working hypothesis [13]. However, some fundamental questions remain open, such as how stability and function are related, and how flexibility plays a role in their modulation [14,15]. Advances in the membrane protein folding field are relatively slower. Expression levels for functional membrane proteins are usually low, and many experimental approaches which are routinely employed for globular proteins are very hard to be performed for the membrane-related ones [16,17].

Thermodynamic characterization of membrane protein stability requires the determination of experimental conditions where unfolding can be reversibly achieved [18], and this has been possible for only a few of these proteins. Moreover, experimental conditions that were successfully attempted for one system frequently do not work in another [19]; thus, no systematic approach seems to be possible by now. Furthermore, kinetic studies that have been essential to determine the folding mechanism of globular proteins [20] are very few for membrane proteins, though the landscape is not as obscure as it could be. Beta barrel membrane proteins such as bacterial porins constitute a model system for which thermodynamic and kinetic folding studies have been successfully performed [21]. Typically these proteins can be denatured using chaotropic agents such as guanidine hydrochloride and urea, and refolded in the presence of amphiphiles [21]. Conversely, helical membrane proteins are resistant to chaotropes, but ionic detergents such as SDS have been shown to be efficient denaturants for some of them [22].

Additional complexities appear when working with multidomain membrane proteins. These proteins contain a membrane related domain and one or more water related structured regions. This very interesting group of proteins is involved in important cellular functions such as active transport, and signal transduction. The ion transport P-type ATPases are typical exponents of this group [23,24]. Their structure is characterized by a membrane-associated domain, which contains the molecular machinery for ion transport, and several soluble domains including a catalytic domain, regulatory sites and, in some cases, soluble metal binding sites. The synchronized interaction among these domains determines the coupling between the catalytic activity and protein interactions with membrane lipids [25] and the transport function [26]. As can be expected, these complexities are translated to difficulties in their study, so that the reported studies on their folding and stability are very few.

## 2. Basic Strategies for Studying Protein Folding and Stability

The stability of proteins can be defined as the resistance of native folded conformations against their disruption by environmental factors conducive to denatured states. Experimental studies on protein folding and stability require three main preconditions: the first one is folding needs to be a reversible process, so that the protein can be unfolded and refolded at equilibrium; the second is to have an agent capable of perturbing the equilibrium, thus modifying the concentrations of native, intermediate, and unfolded species; and the third one is the existence of a measurable signal that monitors the perturbation. Given these three conditions, equilibrium thermodynamics provides the framework for determining the Gibbs free energy change for transfer of the protein from an ideal one molar solution of the pure folded protein in water to an ideal one molar aqueous solution of the pure unfolded protein *i.e.*, ($\Delta G^o_{w\ N\rightarrow U}$), which is the accepted form to quantify the structural thermodynamic stability of a protein [27–29].

Two main strategies are used for disturbing the folding equilibrium of a protein under study: temperature perturbation and the use of chemical denaturing agents.

Temperature extensively modulates the Gibbs free energy by tuning its enthalpic and entropic contributions [29]:

$$\Delta G^o_w = \Delta H^o_w - T \cdot \Delta S^o_w - \Delta Cp^o_w \cdot \left[ (T_o - T) + T \cdot \ln\left(\frac{T}{T_o}\right) \right] \qquad (1)$$

As can be observed in Figure 1A, increasing or decreasing the temperature after reaching a maximal stability produces a decrease in $\Delta G^o_{w\ N\rightarrow U}$, which becomes negative for temperatures higher than the so-called thermal denaturation midpoint ($T_m$) or lower than the cold denaturation temperature ($T_c$). Convexity in Figure 1A is given by the heat capacity change upon unfolding ($\Delta Cp^o_w$), which reports the difference between solvent interactions in the folded and unfolded states, and it has been found to correlate with the change in the solvent-accessible surface area ($\Delta ASA$) as a protein unfolds [30].

On the other hand, chemical denaturants are a group of molecules that stabilize expanded conformations of the polypeptide chain favoring partially folded and unfolded states, thus decreasing the unfolding free energy. The dependence of $\Delta G^o_{w\ N\rightarrow U}$ on the denaturant concentration was empirically explored for many small globular proteins (Figure 1B), typically finding a linear relationship [31]:

$$\Delta G^o = \Delta G^o_w - m_{nu} \cdot [\text{denaturant}] \qquad (2)$$

The coefficient $m_{nu}$, *i.e.*, the derivative $\Delta G^o$ with respect to the denaturant concentration, gives a measure of the cooperative effect of disruption of the protein native structure by the denaturant, and depends on the total $\Delta ASA$ and on the composition of the surface that is exposed upon the unfolding process [30,32]. As can be observed in Figure 1B, ΔG° becomes negative for denaturant concentrations higher than the so-called mid transition concentration ($C_m$).

**Figure 1.** Dependence of the protein folding equilibrium on temperature and the presence of denaturants. (**A**) Temperatures modify $\Delta G^\circ_{w\,N\to U}$ reaching a maximum value which corresponds to the maximal protein stability. Heating or cooling the sample will lead to a decrease in $\Delta G^\circ_{w\,N\to U}$ which will equal zero at the temperature of cold unfolding ($T_c$) and at the melting temperature ($T_m$). At both temperatures half the protein population is folded and the other half is unfolded; (**B**) Protein stability decrease following a linear function of the concentration of the chaotropic agent. $\Delta G^\circ_{w\,N\to U}$ reaches the zero value at the mid-denaturant concentration ($C_m$) where 50% of the population is folded and the other 50% is unfolded.

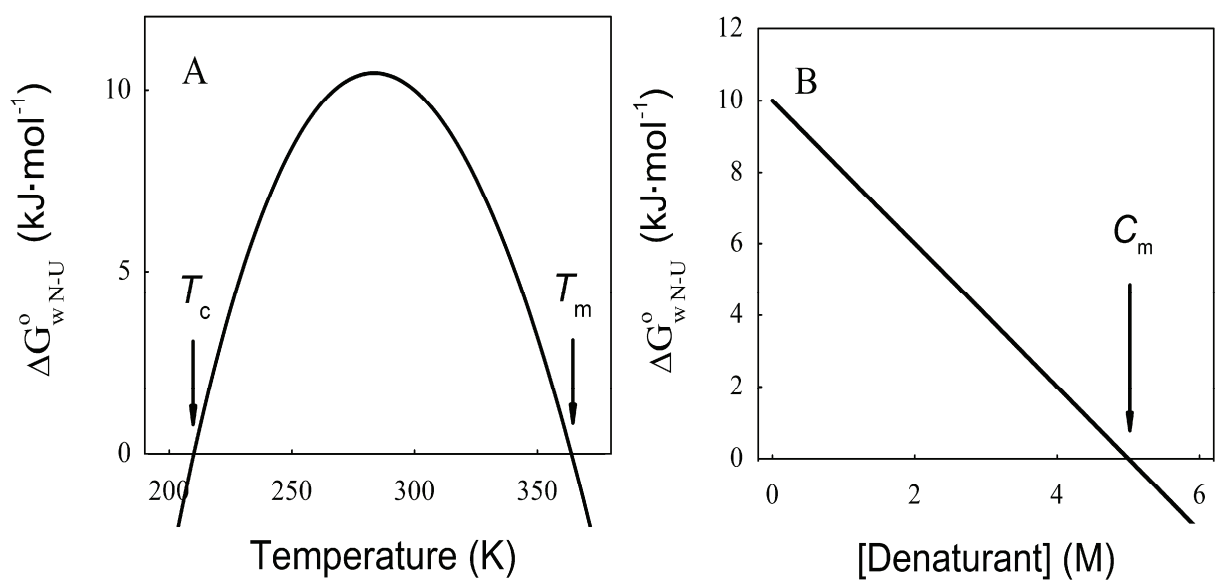

Among various chemical denaturant agents, urea and guanidine hydrochloride (GndHCl) have proven particularly successful. The mechanism by which these compounds denature proteins has been a controversial issue, but recent computational and experimental studies point towards an emerging consensus [33]. These species preferentially accumulate in the water of hydration relative to bulk water, paralleling the Hofmeister chaotropic series of non-Coulombic effects of salt ions on protein processes [34]. It has been proposed that they bind with higher affinity to the unfolded state, solvating charged residues or engaging in hydrogen bonds with charged groups and backbone carbonyl groups [35]. More favorable protein-chaotrope interactions can be formed in the unfolded ensemble than in the native one, thus providing an enthalpic driving force for unfolding [33]. Chaotropes also increase the water solubility of non-polar groups of the backbone and side-chains, thus determining a favorable free energy of transfer for polar residues from the protein core to the aqueous environment. Therefore, the folding equilibrium is shifted toward the unfolded form, which occurs at very high denaturant concentrations (in the order of molar concentrations). The chaotrope-induced denatured state was first described as a random coil chain [28,29]. However, recent studies demonstrate that a statistical ensemble of multiple conformations with a very low content of secondary and tertiary structures better characterizes this state. Indeed, the presence of long-range contacts in this ensemble has been reported by NMR even in the presence of high denaturant concentrations [36].

Another class of denaturant agents corresponds to ionic surfactants, with sodium dodecyl sulfate (SDS) the most representative member. As chaotropes, SDS binds to both native and denatured proteins, and unfolding is driven by the higher affinity for the denatured state. However, the binding mechanism is different [37]. First, the driving force contains an important favorable entropic contribution [38], making the denaturing effect evident at millimolar concentrations, *i.e.*, they are about 1000 times more efficient denaturants than chaotropes. Besides, hydrophobic clusters of SDS molecules are also able to induce helical structures in amphipathic non-structured chains [39]. Thus, the final denatured state has high content of secondary structure. The global shape of protein-SDS complexes was first described by Reynolds and Tanford as rod-like prolate elipsoids [40]. Recently, these complexes were modeled using data from NMR, light scattering, fluorescence spectroscopy and small angle X-ray scattering. Two main spatial arrangements emerge from these studies (Figure 2): the first one, denoted as the decorated micelles model, shows the unfolded protein wrapped around detergent micelles [41–43], whereas the second model, called the pearl necklace or necklace-and-beads model, display micelles on several parts of the unfolded protein chain, randomly decorating the polypeptide backbone [44,45].

**Figure 2.** Structural models for protein-surfactant complexes. (**A**) Surfactant molecules cover the denatured protein chain forming a rod-like micelle; (**B**) Surfactant micelles are decorated by the denatured protein; (**C**) Micelles are formed along the denatured protein forming a pearl-necklace like structure.

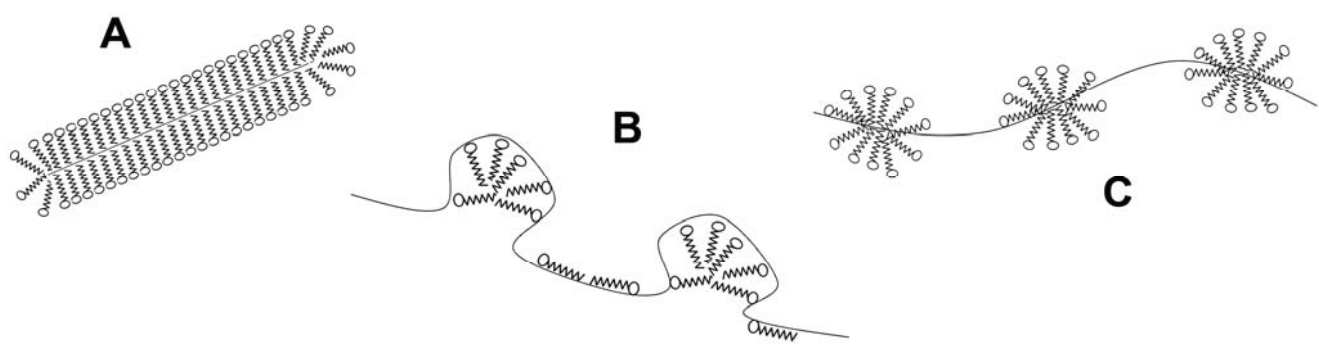

## 3. Temperature Induced Membrane Protein Denaturation

Most membrane proteins undergo spontaneous irreversible thermal denaturation even under mild experimental conditions ([22,46,47], and references therein). Irreversible inactivation has been extensively studied for soluble proteins, and in many cases it is due to either covalent modifications (side-chain oxidation, hydrolysis of peptide bonds, *etc*.) or non-covalent changes leading to aggregation [48–50]. Characterization of the temperature-induced denatured state of diacylglycerol kinase suggests that membrane protein thermal inactivation is not associated with these phenomena, and it was proposed that other non-covalent changes such as the formation of kinetically trapped conformations [51] could be involved.

Irreversibility impedes quantitative thermodynamical analysis of structural stability. In these cases, protein stability can be evaluated by quantifying the time persistence of the native folded conformation (kinetic stability). This irreversible process can be monitored through a measurable signal such as an enzymatic activity, the molar ellipticity in the far UV (indicative of secondary structure) [52] or the intensity of emitted fluorescence either from the intrinsic Trp residues [53] or given by an external fluorescent probe such as 1-anilinenaphthalene-8-sulphonate which reports on hydrophobic regions in protein structure including the transmembrane domain of membrane proteins [54]. The time course of these signals provides a measure of the kinetic stability of the protein.

The copper(I) transport ATPase from *Archaeoglobus fulgidus* [55], the plasma membrane calcium pump [56] and the Na,K ATPase [57] are typical examples where irreversible inactivation was followed by both structural and functional signals. In all these cases, the decrease in enzyme activity is well described by an exponential function suggesting a two-state process involving only fully active and inactive molecules.

$$M = M_o \cdot e^{-k_{obs} \cdot t} \tag{3}$$

where $M_o$ and $M$ are the values of the measurable signal before and after a given preincubation time $t$, and $k_{obs}$ is the observed kinetic coefficient for the irreversible transition.

The decrease in enzyme activity is paralleled by large changes in secondary and tertiary structure. However, the denatured state still conserves important amounts of residual structure, mainly corresponding to the membrane associated regions of the protein [55,58]. Notably, the kinetic coefficients for the structural changes are not statistically different from that for the inactivation process, supporting the idea that enzyme inactivation is associated with partial unfolding and exposition of hydrophobic regions to the solvent. The temperature dependence of the inactivation coefficients is similar for many membrane proteins despite that they are mesophilic or thermophilic, with activation energies in the order of 200–300 kJ·mol$^{-1}$ [55–57,59,60].

On the other hand, thermal stability of membrane proteins reconstituted in mixed micelles of phospolipids and detergents increases when the mole fraction of phospholipids is increased [56]. Maximal stability was found to be dependent of the type of phospholipids included in the micelles, and was also affected when phospholipids undergo chemical modification such as non-enzymatic glycation of phosphatidylethanolamine, a common alteration in uncontrolled *Diabetes mellitus* [61]. The modulation of membrane protein stability by phospholipids can be understood assuming that these proteins have their hydrophobic transmembrane surface covered by a monolayer of amphiphiles in rapid exchange with the bulk amphiphiles [62], and that thermal stability is determined by the composition of this monolayer. When the immobilized boundary layer is predominantly composed of phospholipids, the inactivation rate coefficient reaches its minimum and therefore, the protein thermal stability is maximal. For all the systems assayed, this maximal stability was attained when about 80% of the transmembrane protein surface is covered by phospholipid molecules [56].

## 4. Solvent Denaturation of Membrane Proteins

### *4.1. The Two-State Model, a Paradigm for Small Globular Protein Folding*

The simplest protein unfolding process can be described as an elemental chemical reaction where the reactant is the native protein (N) and the product is the unfolded protein (U):

$$N \underset{k_f}{\overset{k_u}{\rightleftharpoons}} U \tag{4}$$

being $k_f$ and $k_u$ the folding and unfolding rate constants, respectively. Changes in the concentration of $U$ are given by:

$$\frac{d[U]}{dt} = k_u \cdot [N] - k_f \cdot [U] \tag{5}$$

and, at the equilibrium:

$$\frac{[U]_{eq}}{[N]_{eq}} = \frac{f_U}{f_N} = \frac{k_u}{k_f} = K_{eq} = e^{-\frac{\Delta G^o_w - m \cdot [\text{denaturant}]}{R \cdot T}} \tag{6}$$

where $f_N$ and $f_U$ are the mole fraction of the native and unfolded protein at equilibrium, which depend on the denaturant concentration (Figure 3A). Considering that the sum of [$N$] and [$U$] yields the total protein concentration, Equation (5) can be integrated giving

$$[U] = \left([U]_o - [U]_{eq}\right) \cdot e^{-(k_f + k_u) \cdot t} + [U]_{eq} \tag{7}$$

$$[N] = \left([N]_o - [N]_{eq}\right) \cdot e^{-(k_f + k_u) \cdot t} + [N]_{eq} \tag{8}$$

Equations (7) and (8) show that during folding experiments (*i.e.*, when $[U]_o > [U]_{eq}$) the time course of [$N$] and [$U$] follows single exponential curves (decreasing exponentials for $U$ and increasing exponentials for $N$). It can also be observed that the exponential coefficient is the same in both equations and corresponds to the observed kinetic coefficient ($k_{obs}$). When denaturing a protein, these folding and unfolding rate constants are dependent on the denaturant concentration, and empirical exponential relationships have been proposed for accounting these dependences [28].

$$k_{obs} = k_f + k_u = k_f^o \cdot e^{-m_f \cdot [D]} + k_u^o \cdot e^{m_u \cdot [D]} \tag{9}$$

The folding and unfolding kinetic coefficients in water ($k_f°$ and $k_u°$) can be determined by fitting Equation (9) to experimental values of $k_{obs}$ measured in unfolding and refolding experiments performed in the presence of different concentrations of denaturant. Figure 3B shows a typical representation of folding kinetic data in the form of the so-called Chevron Plot. The activation free energy for unfolding ($\Delta G^{\ddagger}$) which correlates with the probability of crossing the energy barrier separating the native from the unfolded state, can be calculated using the transition state theory. In addition, the parameters $m_f$ and $m_u$ in Equation (9) provide information about the surface exposition upon denaturation with respect to the exposed surface in the transition state [30].

Experimentally measured values of $k_f°$ for single domain globular proteins give characteristic times (*i.e.*, the inverse of the $k$) ranging from microseconds to hours. Protein folding rates were theoretically shown to decrease with the size of the protein, and to increase with protein stability ($\Delta G^o_{w\ N \to U}$). When represented as a function of a combined size-stability variable, they fall within a narrow "golden triangle", which also predicts a maximal allowed size for a protein domain that folds under thermodynamic control of about 500 residues [63,64].

**Figure 3.** Equilibrium unfolding curve and *Chevron Plot* for two-state protein unfolding. (**A**) Representation of the equilibrium folded fraction of a protein as a function of denaturant concentration. It can be observed that either at low or high denaturant concentrations, the folding equilibrium is minimally modified by the addition of denaturant. Conversely, at intermediate denaturant concentrations the equilibrium is significantly shifted, leading to a broader change in folded and unfolded concentrations. The maximal slope is related to the cooperativity of the process (*m* value). The concentration where 50% of the population is folded and 50% unfolded is called the mid-denaturant concentration ($C_m$); (**B**) The natural logarithm of the observed kinetic coefficient is plotted as a function of denaturant concentration (*Chevron Plot*). The folding arm is usually explored in refolding experiments where the unfolded protein in the presence of high denaturant concentration is diluted with buffer without denaturant, and the unfolding arm is mainly obtained from kinetic unfolding experiments. This is because determining kinetic coefficients requires a significant change in the signal used to monitor the unfolding/refolding process.

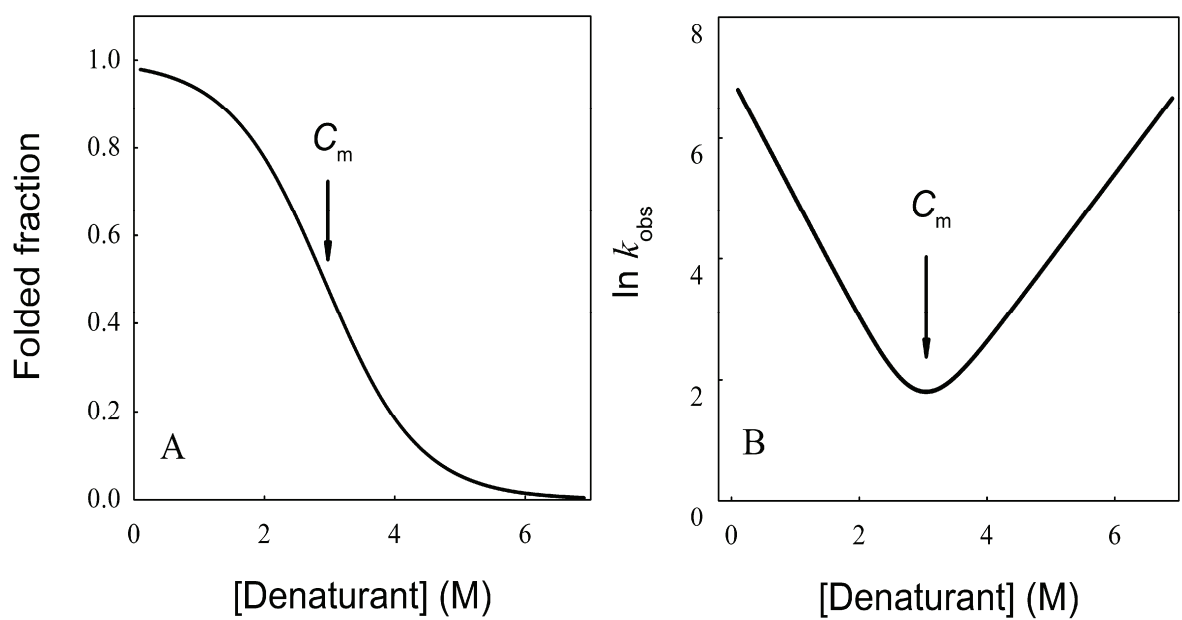

As an example, we will mention the case of the SH3 modules, which form part of many signal transduction and cytoskeletal proteins, and mediates a myriad of protein-protein interactions [65]. SH3 domains possess a distinctive fold (Figure 4) composed of two 3-stranded β-sheets packed orthogonally against each other forming a single hydrophobic core [66]. Besides their biological importance, their ability to independently fold, their small size and the lack of disulfide bonds and cofactors, the variability in the loop regions, and their simple production and purification, make the SH3 domains an attractive model system for testing folding mechanisms [67]. These domains constitute a paradigm for equilibrium folding, emerging as a prototypical two-state folder [68].

**Figure 4.** Structure of the *N*-terminal Src homology 3 domain (Protein Data Bank ID 1CKA). Beta sheets are represented by arrows and protein loops are shown as thin lines.

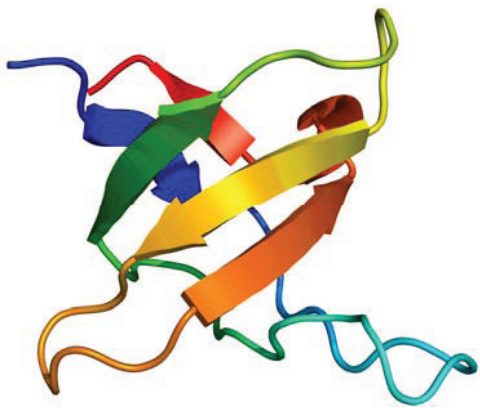

Folding of the src SH3 domain was explored in many buffer systems and ionic strength conditions using temperature and guanidine hydrochloride for perturbing the folding equilibrium [67]. In all these cases, unfolding of src SH3 domains follow a typical two-state mechanism: thermodynamic unfolding parameters in water at 295K were identical for temperature and GndHCl-induced unfolding being $\Delta G^\circ_{w\ N\to U} = 17$ kJ·mol$^{-1}$ and $m_{nu} = 6.7$ kJ·mol$^{-1}$·M$^{-1}$. In addition, folding and unfolding kinetics can be well described by single exponential functions, indicating the lack of intermediates in the folding reaction. The dependence of ln $k_{obs}$ on [GndHCl] has the typical V-shape corresponding to a two-state process with $k_f^\circ = 56.7$ s$^{-1}$, $k_u^\circ = 0.1$ s$^{-1}$, $m_f = 4.2$ kJ·mol$^{-1}$·M$^{-1}$ and $m_u = 1.9$ kJ·mol$^{-1}$·M$^{-1}$. From these data it is possible to calculate an equilibrium $\Delta G^\circ_{w\ N\to U}$ using Equation (6), obtaining a value of 16 kJ·mol$^{-1}$, very close to the thermodynamically determined value. All these data allow the inclusion of the src SH3 domain on the list of single-domain proteins that fold without detectable populations of partially folded intermediates. Furthermore, it can be demonstrated that the ratio between $m_f$ and $m_{nu}$ gives an idea of the surface change between the transition state and the folded ensemble of conformations. In this case, $m_f/m_{nu} = 0.6$ suggesting that about two thirds of the buried surface area of the folded protein is excluded from the solvent in the transition state [67].

*4.2. Membrane Protein Folding in Solution*

4.2.1. Beta Barrel Bacterial Outer Membrane Proteins

The outer membrane proteins (OMPs) of Gram-negative bacteria, mitochondria, and chloroplasts constitute a diverse group of β-barrel proteins that include adhesins, architectural proteins, passive diffusion pores, siderophore receptors, efflux channels, protein translocation pores, and some membrane enzymes such as lipases, proteases, and palmitoyl transferases [69].

Despite that they constitute a very small fraction of the membrane proteins codified in the known genomes [70], β-barrel membrane proteins are very well represented in the Protein Data Bank with almost 190 hits.

OMPs share a very simple structural framework consisting of 8 to 24 β-strands embedded in the membrane oriented in an antiparallel manner [71], and may also have extramembrane domains and loops connecting the individual β-strands (Figure 5). Some OMPs oligomerize to form trimeric porin channels [72,73]. Transmembrane strands consist of alternating hydrophobic and hydrophilic residues, thus reducing its overall hydrophobicity and making possible their solubilization by chemical denaturants [47]. In addition most of them can be refolded under equilibrium conditions in the presence of phospholipid vesicles [19,74] or surfactant micelles [75] making it possible to perform folding-function-stability studies [76].

**Figure 5.** Structure of the Outer membrane protein A (OmpA) from *E. coli* (Protein Data Bank ID 1G90). Beta sheets are represented by arrows and protein loops are shown as thin lines. The beige area represents the hydrophobic core of the membrane, while the light-indigo regions refer to the hydrophylic head groups of phospholipids.

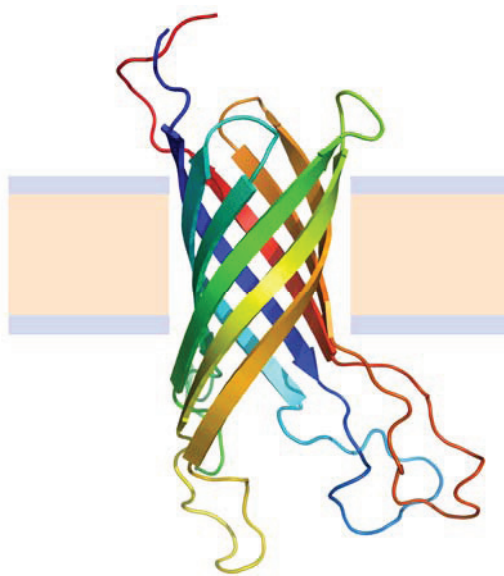

Since the first successful refolding experiment of an OMP into DMPC vesicles [74] many efforts have been performed in deciphering porins folding mechanism. It was determined that OMPs refolding efficiency in vesicles depends on membrane thickness, phospholipid charge, acyl chain saturation, vesicle size and curvature, temperature, ionic strength, pH and the denaturing agent used [21]. Unfortunately, optimal conditions are particular for each OMP [19].

Three transitions along the folding pathway of OmpA were distinguished. A water-soluble collapsed state, consisting of a mix of random coil, α-helix and β-sheet, has been identified as an early folding intermediate state [77]. This collapsed state associates to lipid vesicles adopting a partially folded β-sheet structure, which in a slow process generates the full inserted folded form [78]. Membrane associated folding intermediates were elegantly characterized by Kleinschmidt *et al*. [79,80] using lipid vesicles prepared with a set of brominated phospholipids, which are known to quench the intrinsic Trp fluorescence [81], and single-Trp mutants of OmpA. Experiments were designed combining equilibrium and kinetic approaches. Equilibrium experiments were performed by measuring Trp fluorescence of the wild type and the single Trp mutants to test that the reversibility requirement was fulfilled. Unfolding in 8 M urea proceeds with a significant decrease in total Trp fluorescence intensity and a red-shift in the center of the mass of the fluorescence spectra. Dilution of this system into a lipid-rich media without denaturant results in the recovery of the fluorescence spectral properties, indicating that OmpA is efficiently refolded. The authors registered the time course of fluorescence quenching at different temperatures, identifying three membrane-associated folding intermediates at different distances from the center of the bilayer. Detection of folding intermediates was temperature-dependent so that it was possible to isolate each one. The first intermediate was found to be adsorbed to the surface without a high degree of structure, whereas the following ones have increasing quantities of β-sheet structure and a deeper penetration into the membrane. The study at different temperatures allowed exploration of the folding landscape, and investigation of OmpA folding-insertion into biological bilayers at different levels of complexity. Following this work, the authors propose a detailed multistage concerted folding model for this β-barrel protein in which all four β-hairpins of OmpA insert and translocate across the membrane by a synchronous concerted mechanism. The proposed folding-insertion mechanism is clearly different from that proposed for helical membrane proteins by Popot and Engelman [82]. In that model, transmembrane helices form independent transmembrane folding units (stage I) and they then assemble to form the native three-dimensional structure (stage II). The demonstration that a general folding paradigm can not be posited for membrane proteins and the analysis of the folding-insertion process in terms of a folding landscape, make this work pioneering.

Although it has been well established that insertion is coupled to native structure consolidation, lipid composition of the membranes plays important roles in determining the folding pathway. It was demonstrated that folding kinetics are very sensitive to membrane properties such as membrane thickness [83]. Furthermore, Hong *et al*. [84] found that the lateral pressure that membranes exert on the protein produces a tremendous effect on protein stability. As the hydrophobic surface of the protein is covered with lipids, a high lateral pressure is exerted onto the protein. In these conditions the native state would be more stabilized than the possible intermediates, and thus the folding process will follow a simple two-state mechanism. Conversely, as the membrane becomes thinner, the lateral pressure decreases and water molecules can penetrate the membrane and interact with the protein. Then, intermediate structures are more prone to be stabilized and the folding route increases in complexity.

In another interesting study Debnath *et al*. [85] showed that production of *in vitro* complementary sequences of OmpA could yield a fully folded protein. They engineered a variant of OmpA with cleavage sites, and then digested the protein into complementary fragments. These fragments were unfolded with urea and refolding was assayed, demonstrating that 25%–35% of the fragment populations can associate producing a native like protein. This result demonstrates that although membrane insertion and folding may be a cooperative process, the fragments can associate *in vitro* without any additional components.

4.2.2. Helical Membrane Proteins

The main group of membrane proteins correspond to helical membrane proteins which are found in all cellular membranes, are involved in diverse and critical cellular functions, and include nearly all medically relevant membrane proteins. These proteins are composed of one or several α-helices inserted into biological membranes. The structure of bacteriorhodopsin (Figure 6)—a classical model for helical membrane proteins—showed a tightly packed bundle of typical α-helices in the membrane [86]. However, recent high-resolution structures show that transmembrane helices contain frequent bends and kinks, some break in the middle and change directions, and others begin and end within the core of the membrane [87].

**Figure 6.** Structure of Bacteriorhodopsin from *Halobacterium salinarum* (Protein Data Bank ID 2BRD). Alpha helices are represented by coiled ribbons, and protein loops are shown as thin lines. The beige area represents the hydrophobic core of the membrane, while the light-indigo regions refer to the hydrophilic head groups of phospholipids.

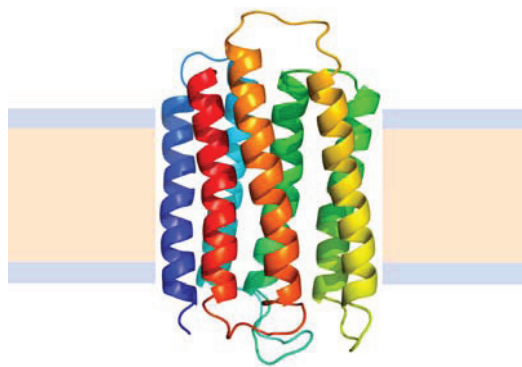

To date, there are only five members of this group for which a reversible unfolding has been possible: bacteriorhodopsin [88,89], diacylglycerol kinase [90], the potassium channel KcsA [91], the disulfide bond reducing protein DsbB [92] and the copper transport ATPase from *Archaeoglobus fulgidus* [93]. This limited information results in a poor knowledge about the mechanisms of insertion and folding.

Helical membrane proteins appear resistant to chemical denaturation by urea or guanidine hydrochloride [94] or, in some instances, treatment with these denaturants leads to irreversible unfolding [95]. The inefficiency of chaotropes to unfold helical membrane proteins has been linked to the structural organization of the transmembrane segments as α-helix bundles. This retains a moderately hydrophobic core by exposing highly hydrophobic residues to the membrane lipids thus preventing the access of polar chaotropes [88]. However, ionic detergents such as SDS have been efficiently used to unfold some of these proteins under equilibrium conditions [22,96]. The amphipathic nature of these detergents combine the possibility of disrupting tertiary contacts and solubilizing the secondary structure elements buried in biological membranes.

Early works of Lau and Bowie with diacylglycerol kinase [90] showed a multiphasic unfolding curve upon increasing the concentration of sodium dodecyl sufate. At very low SDS concentrations the oligomeric structure of the native protein is disrupted, and two unfolding transitions were observed when increasing SDS concentration. To analyse this result, the authors constructed a set of point mutants of single Trp residues distributed within the membrane interacting region and the cytoplasmic domain. Mutants containing "cytoplasmic" Trp residues showed only the first unfolding transition when monitored by intrinsic fluorescence, while the single Trp mutants in the transmembrane helices only displayed the second transition. Thus, the authors conclude that after disruption of the oligomeric structure the cytoplasmic region of diacylglycerol kinase monomers was first unfolded, followed by the denaturation of the transmembrane domain.

Booth and coworkers [97] have extensively studied the unfolding of bacteriorhodopsin induced by SDS making significant contributions to the membrane protein folding field. The retinal chromophore bound to the native protein by a Schiff base permits testing in the functional folded state. It was demonstrated that bacteriorhodopsin can be efficiently unfolded by SDS, but refolding in this detergent recovered a protein conformation with 56% of its native secondary structure and minimal tertiary structure, without the ability to bind its cofactor and being nonfunctional [98,99]. Reversible refolding to a fully functional form was possible with the cofactor and either non-denaturing detergents, lipids in mixed micelles or lipid vesicles in the refolding medium, The unfolding equilibrium curve of bacteriorhodopsin determined in these conditions was well described by a reversible two-state model with $\Delta G^\circ_{w\,N\to U}$ of about 24 kJ·mol$^{-1}$. However, the unfolding kinetics showed that the folding mechanism is indeed more complex including several intermediates, and that the pathways for unfolding and refolding are different [100]. The relation between the folding and unfolding rate constants and SDS concentration suggest a linear free energy relationship similar to that described for small globular proteins. Besides, the structure of the unfolding transition state was characterized using a set of single mutants, being closer to the SDS-unfolded state than to the native one. This behavior is clearly different from that described for most small globular proteins where the structure of the transition state is closer to that of the native state [101]. On the other hand kinetic studies suggested that the folding route is polarized: while SDS disrupts the helical packing of the outer helices, some of the internal ones are resistant, so that the protein conserves a large amount of secondary structure in the denatured state [102]. Further attempts to characterize this unfolded state showed that although the secondary structure contains significant amounts of native signatures, the tertiary structure was mainly disrupted [103]. These important differences with the unfolded state of soluble proteins open interesting questions about how to compare unfolding processes when the final unfolded states are very different.

A particular, and exceptional case of alpha helical membrane proteins are the multidomain group (Figure 7); these proteins contain, besides the transmembrane region, soluble domains that are very similar to globular soluble proteins [104], and, in fact, look like many globular proteins attached to transmembrane helices. Studying their folding is particularly complex and there are very few proteins in this group for which stability studies were successful; among them P-type ATPases constitute an interesting example, because, in addition, several crystal structures of members of this membrane protein family have been deposited in the Protein Data Bank during the last years [105,106].

**Figure 7.** Structure of the Cooper transport ATPase from *Legionella pneumophila* (Protein Data Bank ID 3RFU). Alpha helices are represented by coiled ribbons, beta sheets by arrows and protein loops are shown as thin lines. The beige area represents the hydrophobic core of the membrane, while the light-indigo regions refer to the hydrophilic head groups of phospholipids.

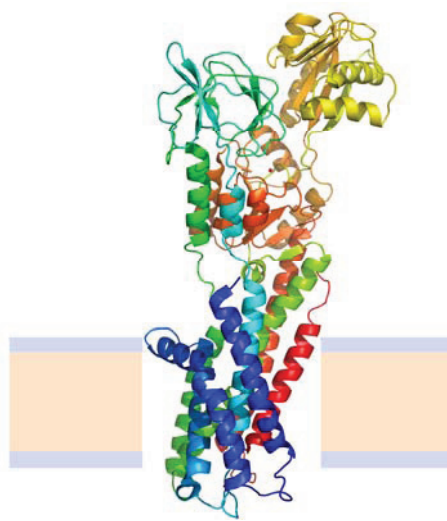

The Cu(I) transport ATPase from the hyperthermophile *Archaeoglobus fulgidus* (CopA) reconstituted in mixed micelles is one of the cases where reversible unfolding was successful [93]. This protein could be denatured with guanidine hydrochloride losing most of its secondary and tertiary structure as seen by far and near UV circular dichroism. In addition, hydrophobic patches in CopA, mainly located in the transmembrane region, were disrupted as indicated by 1-anilino-naphtalene-8-sulfonate fluorescence. Nevertheless, the unfolded state had a small but detectable amount of residual structure, which might play a key role in both CopA folding and adaptation for working at high temperatures. Refolding of CopA from this unfolded state led to recovery of full biological activity and all the structural features of the native enzyme.

CopA unfolding showed typical characteristics of a two state process with $\Delta G_w^\circ$ = 12.9 kJ·mol$^{-1}$, $m_{nu}$ = 4.1 kJ·mol$^{-1}$·M$^{-1}$ and $\Delta C p_w^\circ$ = 0.93 kJ·mol$^{-1}$·K$^{-1}$. These values were relatively lower than those expected for mesophilic proteins of similar molecular mass that unfold according to a two-state model. An approximately linear dependence of $\Delta G_w^\circ$ on the number of protein residues is well established for the unfolding of small soluble monomeric proteins (smaller than 200 residues) [107]. Booth and Curnow have shown that bacteriorhodopsin, diacylglycerol kinase and the potassium channel KcsA, seem to fit within this trend [107]. On the contrary, CopA $\Delta G_w^\circ$ does not follow this tendency. In this way, it is worth mentioning that there is a small number of large proteins (none of them a membrane protein), for which thermodynamic stability was assessed and only in a few cases they unfold following a two-state process. This set of large soluble proteins also show unusually low values of $\Delta G_w^\circ$, e.g., human serum albumin for which $\Delta G_w^\circ$ is a quarter of the value expected according its molecular mass [108].

On the other hand, the *A fulgidus* CopA structure shows five distinct structural regions, including the transmembrane α-helices and soluble catalytic and regulatory domains [105,109]; then, the description of its unfolding as a two-state process does not seem to be appropriate . For instance, the N→U transition is likely to include an ensemble of unfolding intermediate states with closely related conformations indistinguishable by far UV-CD and Trp fluorescence. Moreover, it was demonstrated that the region corresponding to low GndHCl concentrations contains inactive enzymes with an apparent native form as suggested by the spectroscopic characterization. The presence of these non-detectable unfolding intermediates is known to lead to a significant underestimation of both $m_{nu}$ and $\Delta G_w^\circ$ values [110].

Of course, complexities in the study of these proteins have limited the progress in the protein folding area. However, the advances produced during the last years permit the expectation of a substantial increase in thermodynamic and kinetic helical membrane protein folding information in the near future.

## 5. Conclusions

The membrane protein folding field is being increasingly explored in recent years as new structural data is obtained. Approximately 400 unique high-resolution membrane protein structures are now available in the Protein Data Bank. This information is leading to new insights that aid the analysis of structure-function relationships and provides the framework for the interpretation of thermodynamic and kinetic data on the folding of these proteins in membrane-like environments. However, further work in this area is still necessary to attempt to posit some general rules. As was proposed by Finkelstein [18], it may be that future developments will be based on the physics of single domain proteins, in a similar way as chemistry is based on the physics of atoms and electrons.

**Acknowledgments**

This work was supported by ANPCyT PICT2010 01876, UBACyT 20020100100048 Grants.

**Author Contributions**

Ernesto A. Roman and F. Luis González Flecha wrote the review.

**Conflicts of Interest**

The authors declare no conflict of interest.